\documentclass[superscriptaddress,prb,twocolumn]{revtex4}

\usepackage{amsmath}
\usepackage{amssymb}
\usepackage{epsfig}
\usepackage{graphicx}
\usepackage{wasysym}
\usepackage{multirow}

\newcommand \be{\begin{equation}}
\newcommand \ee{\end{equation}}
\newcommand \bes{\begin{equation*}} 
\newcommand \ees{\end{equation*}}
\newcommand \bea{\begin{eqnarray}}
\newcommand \eea{\end{eqnarray}}
\newcommand \beas{\begin{eqnarray*}} 
\newcommand \eeas{\end{eqnarray*}}
\newcommand \bfg{\begin{figure}}
\newcommand \efg{\end{figure}}
\newcommand \bfgs{\begin{figure*}} 
\newcommand \efgs{\end{figure*}}
\newcommand \bwt{\begin{widetext}}
\newcommand \ewt{\end{widetext}}

\newcommand \nd{{\vphantom{\dagger}}} 
\newcommand \bra{\langle}
\newcommand \ket{\rangle}

\newcommand \Sec[1]{Section~\ref{#1}}
\newcommand \App[1]{Appendix~\ref{#1}}
\newcommand \Equ[1]{Eq.~(\ref{#1})}

\newcommand{\Fig}[1]{Fig.~\ref{#1}}

\newcommand \etal{{\it et al.}}

\newcommand \mirror{\sigma}
\newcommand \SGid{\mathbf{1}}
\newcommand \phase{\phi}

\newcommand \im{{i}}
\newcommand \dif{\mathrm{d}}
\newcommand \deriv{D}

\newcommand \veca{\mathbf{a}}
\newcommand \vece{\mathbf{e}}
\newcommand \vecd{\mathbf{d}}

\newcommand \vecS{\mathbf{S}}
\newcommand \vecr{\mathbf{r}}
\newcommand \vk{\mathbf{k}}

\newcommand \vsigma{\mbox{\boldmath$\sigma$}}
\newcommand \vm{\mathbf{M}}
\newcommand \us{\uparrow}
\newcommand \ds{\downarrow}

\begin{document}

\title{Schwinger Boson Mean Field Theories of Spin Liquid States on 
Honeycomb Lattice: Projective Symmetry Group Analysis
and Critical Field Theory.}
\author{Fa Wang}
\affiliation{Department of Physics, Massachusetts Institute of Technology, 
Cambridge, MA 02139}

\begin{abstract}
Motivated by the recent numerical evidence\cite{QMC} of 
a short-range resonating valence bond state in 
the honeycomb lattice Hubbard model, 
we consider Schwinger boson mean field theories of possible spin liquid states 
on honeycomb lattice. From general stability considerations 
the possible spin liquids will have gapped spinons coupled to 
Z$_2$ gauge field. We apply the projective symmetry group(PSG) method 
to classify possible Z$_2$ spin liquid states within 
this formalism on honeycomb lattice. It is found that there are only 
two relevant Z$_2$ states, differed by the value of gauge flux, 
zero or $\pi$, in the elementary hexagon. The zero-flux state is 
a promising candidate for the observed spin liquid and 
continuous phase transition into commensurate N\'eel order. 
We also derive the critical field theory for this transition, 
which is the well-studied O(4) invariant 
theory\cite{ChubukovPRL,ChubukovNPB,Isakov}, and 
has an irrelevant coupling between Higgs and boson fields 
with cubic power of spatial derivatives as required by lattice symmetry. 
This is in sharp contrast to the conventional theory\cite{SachdevRead}, 
where such transition generically leads to 
incommensurate magnetic order. In this scenario the Z$_2$ spin liquid 
could be close to a tricritical point. Soft boson modes will exist at 
seven different wave vectors. This will show up as low frequency 
dynamical spin susceptibility peaks not only at the $\Gamma$ point 
(the N\'eel order wave vector) but also at Brillouin zone edge center 
$M$ points and twelve other points. Some simple properties of 
the $\pi$-flux state are studies as well. 
Symmetry allowed further neighbor mean field ansatz are derived 
in Appendix which can be used in future theoretical works along this direction. 

\end{abstract}
\pacs{}
\maketitle

\tableofcontents

\section{Introduction}\label{sec:intro}
Quantum ground state of a spin system without any spontaneous symmetry 
breaking, the so-called spin liquid, in two or higher spatial dimensions, 
has been a subject of intense research since it was first proposed 
more than thirty years ago\cite{Anderson,FazekasAnderson}. 
These states, sometimes called resonating valence bond(RVB) states, 
generically appear in two varieties, 
the ``short-range RVB state'' with a gap to spin-carrying excitations, 
and the ``critical spin liquid'' with gapless spin excitations. 
Recently several candidate materials\cite{CsCuCl, kappaET,herbertsmithite} 
have emerged for spin liquids in two spatial dimensions(2D). 
Interestingly they all have gapless spin excitations. 
Many parent Hamiltonians have also been constructed for spin liquids in 
2D\cite{ChayesKivelson,Batista,RamanSondhi,Thomale}. 
However it remains unclear theoretically whether a simple and natural spin 
Hamiltonian, e.g. the Heisenberg model, can have a spin liquid ground state on 
some 2D lattices. 
For common bipartite 2D lattices, the square and honeycomb lattices, 
quantum Monte Carlo 
(QMC)\cite{QMCHeisenbergSquare,QMCHeisenbergHoneycomb} 
and other calculations
\cite{SpinWaveSquare,HuseElser,Huse,SpinWave,Singh,SeriesExpansion,Jafari} 
have clearly shown the long-range magnetic order in the ground state of 
the nearest-neighbor Heisenberg 
model. 
Therefore frustration is usually considered as an important ingredient
for stabilizing the putative spin liquid states. 

In an exciting paper by Meng {\etal}\cite{QMC}, 
the half-filled Hubbard model on honeycomb lattice \Equ{equ:Hubbard} 
was carefully studied by quantum Monte Carlo calculations. 
The model simply consists of hopping of electrons on 
nearest-neighbor bonds $<ij>$ 
and onsite repulsion between two spin species labeled by $\alpha=\us,\ds$,
\be
H=-t\sum_{<ij>,\alpha}(c_{i\alpha}^\dagger c_{j\alpha}^\nd
+c_{j\alpha}^\dagger c_{i\alpha}^\nd)+U\sum_{i}n_{i\us}n_{j\ds}.
\label{equ:Hubbard}
\ee
Varying the only parameter in the problem, 
the ratio of onsite repulsion $U>0$ and electron hopping $t$, 
three different phases were observed. 
With small coupling $U/t < 3.5$ the system is a semi-metal 
with Dirac-like dispersion.
For large coupling $4.3 < U/t$ the system develops long range magnetic order. 
In the intermediate coupling region $3.5 < U/t < 4.3$ 
a very interesting state with both single-particle gap and spin gap appears. 
Various symmetry breaking scenarios were checked in this state 
and then ruled out. 
It was thus concluded that this state is a genuine short-range RVB state. 

This is somewhat surprising considering both weak and strong coupling limits.
Starting from the weak coupling limit, 
with the single-particle gap develops continuously 
as observed in the calculation\cite{QMC}, 
it was expected that the spin dynamic will either inherit 
the gapless nature of the small $U$ semi-metal phase\cite{Hermele}, 
or develop certain kind of spontaneous symmetry 
breaking. 

In the strong coupling large $U\to +\infty$ limit 
the low energy Hamiltonian is the nearest-neighbor  
spin-1/2 Heisenberg antiferromagnetic(AFM) model 
whose ground state has long-range colinear 
N\'eel order\cite{QMCHeisenbergHoneycomb} 
and must have gapless spin-wave excitations as Goldstone modes.
Indeed a magnetic order was seen in the strong coupling region $4.3 < U/t$
in the numerical simulation\cite{QMC}. 
Moreover the magnetic order parameter and spin gap 
seem to both vanish continuously at the critical point $U/t\approx 4.3$. 
This raises the hope to understand the observed ``short-range RVB state'', 
at least in the large $U/t$ part of the parameter range, 
by going from the strong coupling side. 
Although the conventional wisdom\cite{SachdevRead,Stability} is 
that such continuous quantum phase transition between 
colinear magnetic order and gapped spin liquid is impossible. 

\begin{figure}
\includegraphics{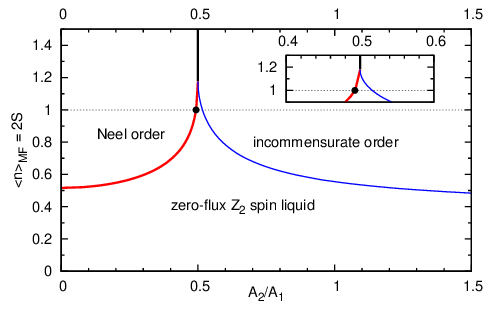}
\caption{(Color online) 
Mean field ``phase diagram'' of the zero-flux state. 
Horizontal axis is the variational parameter, ratio between 
next-nearest-neighbor and nearest-neighbor mean field couplings, $A_2/A_1$. 
Vertical axis is the average boson density $\bra\hat{n}\ket_{\rm MF}$. 
The dash line $\bra\hat{n}\ket_{\rm MF}=1$ indicates 
the boson density of spin-1/2 system. 
Solid lines are phase boundaries. 
The red solid line between the zero-flux Z$_2$ spin liquid and the
N\'eel order is a continuous transition
described by the field theory \Equ{equ:action}. 
The vertical solid black line between the two ordered states 
is a first order transition. 
The blue line between the Z$_2$ spin liquid and the  
incommensurate magnetic order has yet to be studied
but is likely a continuous transition. 
There is a very small parameter range of $0.493<A_2/A_1<0.516$ 
(see also the inset) 
such that a spin-1/2 system will be a gapped Z$_2$ spin liquid, 
which is a promising explanation of the observed spin liquid\cite{QMC}. 
The variational parameter $A_2/A_1$ can in principle be tuned 
by physical parameters. For example, 
as argued in \Sec{sec:fluxstates}, increase of $U/t$ will decrease $A_2/A_1$, 
which can drive a continuous magnetic ordering transition at 
the crossing point (black dot) of the dash line and the red solid line. 
}
\label{fig:kappac-zeroflux}
\end{figure}

In the strong coupling regime, 
with single particle gap much larger than 
the spin gap(zero in magnetic ordered phase), 
it is reasonable to describe the low energy physics by 
an effective spin-1/2 Hamiltonian, 
which can be derived from 
the Hubbard model and should be\cite{MacDonald} (up to $t^4/U^3$ order)
\be
H_{\rm spin}=
\sum_{<ij>} \left (\frac{4t^2}{U}-\frac{16t^4}{U^3}\right )\vecS_i\cdot\vecS_j
+\sum_{<<ij>>} \frac{4t^4}{U^3}\,\vecS_i\cdot\vecS_j+\dots
\label{equ:Hspin}
\ee
where $<<ij>>$ are next-nearest-neighbor bonds. 
As the ``short-range RVB'' region is still close to 
the single-particle gap opening transition(Mott transition), 
the spin Hamiltonian should be much more complex 
than this leading order Heisenberg model, 
i.e. have strong couplings of further neighbors 
and/or four and even more spins. 
Solving the exact spin model will likely not be easier than 
solving the original Hubbard model. 
In this paper we take a different approach. 
Using symmetry analysis we completely classify 
all possible stable gapped spin liquid states within 
the Schwinger boson formalism.
It turns out that there are only two relevant states, differed 
by the gauge invariant flux, zero or $\pi$, in a hexagon. 
Some signatures of these two spin liquid states will be derived  
which may be checked in numerical simulations. 
The zero-flux state turns out to be a very promsing candidate 
for the observed short-range RVB state. 
We obtain a mean field ``phase diagram'' (\Fig{fig:kappac-zeroflux}) for it 
in terms of a variational parameter, which could qualitatively agree with 
the behavior of the Hubbard model close to the magnetic transtion. 
Our symmetry analysis fixes symmetry allowed forms of 
further neighbor mean field couplings, which will be useful
for later theoretical studies of spin liquids on honeycomb lattice.  

The outline of this paper is as follows.
In \Sec{sec:SBMFT} we briefly describe the formalism of  
Schwinger boson mean field theory.
In \Sec{sec:SBPSG} 
we apply the projective symmetry group method developed in
Ref.~\cite{SBPSG} to classify all Z$_2$ Schwinger boson states
on honeycomb lattice. Details of the derivation are presented in 
\App{app:PSG}. 
Two out of 32 possible Z$_2$ states are particularlly relevant here 
and we derive the mean field ansatz up to fourth neighbors in \App{app:ansatz}. 
In \Sec{sec:fluxstates} 
we study some simple properties of 
the two Z$_2$ Schwinger boson states emerged from the PSG analysis. 
And we derive the continuum field theory for the transition from 
the zero-flux Z$_2$ spin liquid to the N\'eel order in 
\App{app:fieldtheory}. 
Conclusions and outlook of further developments are summarized in 
\Sec{sec:conclusion}.

\section{Schwinger Boson Mean Field Theory for Z$_2$ Spin Liquids.}
\label{sec:SBMFT}
A microscopic theory of spin liquid usually involves fractionalized
spin-carrying particles, the spinons, 
which are strongly coupled to certain 
emergent gauge field\cite{ArovasAuerbach,ReadSachdev,SachdevRead,WenSU2}. 
It is generally believed that, when the spinons are gapped, 
the system is stable only 
if the gauge field takes discrete values\cite{ReadSachdev,Stability} 
(some exotic counter-examples exist like the doubled Chern-Simons 
model of Levin and Wen\cite{LevinWen} but will not be considered here). 
The natural candidate of such discrete gauge field for 
short-range RVB state is the Z$_2$(Ising) gauge theory\cite{Sondhi}. 
Thus throughout this paper we will assume a Z$_2$ spin liquid state on
the honeycomb lattice without breaking of any physical symmetry. 

There are several serious problems of the Z$_2$ spin liquid assumption 
in the context of the QMC result\cite{QMC}. 
First if the magnetic ordered phase is continuously connected to 
a Z$_2$ spin liquid, it will usually be non-colinear 
and incommensurate\cite{SachdevRead}, 
unlike the observed commensurate N\'eel-type order. 
However it will be seen later in this paper that 
this expectation is not correct on honeycomb lattice. 
Also it seems 
that the possibility of non-colinear magnetic order has not been carefully 
checked in the paper by Meng {\etal}\cite{QMC}. 
Thus we believe this argument against a Z$_2$ spin liquid explanation 
may be circumvented. 
The second problem is the claim made by Meng {\etal}\cite{QMC} that 
topological degeneracy was not observed, while a Z$_2$ spin liquid on a
torus should have four-fold degenerate ground states. 
But it was acknowledged that their numerical method might have 
missed the degenerate ground states in other topological sectors. 
Despite this uncertainty we believe that it is still meaningful to 
thoroughly study the possibilities of Z$_2$ spin liquids on honeycomb lattice.

Another issue for the Schwinger boson formalism 
is that it is not convenient for the description of 
the seemingly continuous Mott transition around $U/t\approx 3.5$ 
in the numerical results\cite{QMC}. 
We will refrain from considering that parameter range in this paper, 
and strictly limit ourselves in the strong coupling region
with large single particle gap. 

To continuously evolve from a magnetic ordered state 
to a Z$_2$ spin liquid with spin gap, a natural approach
is to decompose each spin into two bosonic spinons, the Schwinger 
bosons\cite{ArovasAuerbach,ReadSachdev,SachdevRead}. 
The magnetic ordering transition then becomes the condensation of 
these bosons\cite{ArovasAuerbach,ReadSachdev,SachdevRead,Sachdev}. 
And a large-$N$ Sp($N$) generalization has been formulated to study 
the problem in a controlled $1/N$ 
expansion\cite{ArovasAuerbach,ReadSachdev,SachdevRead}. 
It is also possible to get a gapped Z$_2$ spin liquid 
from fermionic spinons\cite{WenSU2} but 
that scenario will not be considered in this paper.
In this paper we will not use the Sp($N$) language, but the PSG 
analysis can be directly applied to the large-$N$ theory. 

In the following we briefly recall the formulation of 
the Schwinger boson mean field theory. More details can be found 
in, for example, Ref.~\cite{Sachdev}.

The bosonic representation of spin $\vecS_i$ on site $i$ is
\be
\vecS_i=\frac{1}{2}\sum_{\alpha,\beta}b_{i\alpha}^\dagger
 {\vsigma}_{\alpha\beta} b_{i\beta}^\nd
\label{equ:SchwingerBoson}
\ee
with boson operators $b$, spin indices $\alpha,\beta=\us,\ds$, 
and Pauli matrices $\vsigma$. 
For this to be a faithful representation of the spin system 
a constraint on the total boson number must be imposed,
\be
\hat{n}_i\equiv\sum_{\alpha}b_{i\alpha}^\dagger b_{i\alpha}^\nd = 2S
\label{equ:constraint}
\ee
where $S$ is the size of the spin. 
For spin-1/2 model, $S=1/2$, the boson density should be unity. 
This hard constraint will be relaxed in 
the mean field treatment so it is only satisfied on average under 
the mean field state, 
\be
\bra \hat{n}_i\ket_{\rm MF}=\kappa
\label{equ:averageconstraint}
\ee
where $\bra\cdot\ket_{\rm MF}$ means expectation value in the mean field 
theory, 
and the average boson density $\kappa$ can also be taken as a 
parameter\cite{Sachdev}.

Possible mean field decouplings of Heisenberg interaction 
$\vecS_i\cdot \vecS_j$ 
can be suggested from the operator identities ($i\neq j$)
\be
\begin{split}
& 
\vecS_i\cdot \vecS_j=-2\hat{A}_{ij}^\dagger\hat{A}_{ij}^\nd
+(1/4)\hat{n}_i\hat{n}_j
\\
=
&
-(1/4)\hat{n}_i\hat{n}_j+2 \hat{B}_{ij}^\dagger \hat{B}_{ij}^\nd
=\hat{B}_{ij}^\dagger \hat{B}_{ij}^\nd-\hat{A}_{ij}^\dagger\hat{A}_{ij}^\nd
\end{split}
\ee
where 
$\hat{A}_{ij}=(1/2)(b_{i\us}b_{j\ds}-b_{i\ds}b_{j\us})$ and
$\hat{B}_{ij}=(1/2)(b_{i\us}^\dagger b_{j\us}^\nd +
 b_{i\ds}^\dagger b_{j\ds}^\nd)$
are both SU(2) invariant. 

A mean field theory for Heisenberg AFM model will generally 
include both $\hat{A}$ and $\hat{B}$ terms\cite{Gazza,Lefmann,Misguich},
\be
\begin{split}
H_{\rm MF}=\
&
\sum_{i,j}
(A_{ij}^*\hat{A}_{ij}-B_{ij}^*\hat{B}_{ij}+{H.c.})
+\sum_{i}\mu_i (\hat{n}_i-\kappa)
\\ &
+\sum_{i,j}(A_{ij}^*A_{ij}-B_{ij}^* B_{ij})/J_{ij}
\end{split}
\label{equ:SBMF}
\ee
where $A_{ij}=-A_{ji}$, $B_{ij}=B_{ji}^*$ are complex numbers called 
the mean field ansatz, and
the chemical potential $\mu_i$ is introduced to achieve the average 
constraint \Equ{equ:averageconstraint}. 
For translationally invariant states $\mu_i=\mu$ are uniform. 
And $A_{ij}$($B_{ij}$) on symmetry related bonds will have the same magnitude. 
Both $A$ and $B$ terms have been consistently generalized to 
the theory of Sp($N$) magnets and the mean field Hamiltonian 
can be regarded as a saddle point solution of 
the Sp($N$) action after Hubbard-Stratonovich transformation\cite{Coleman}.
Here we will not use the Sp($N$) language and we will regard 
the mean field theory as a variational approach for 
general spin models even beyond Heisenberg model.

The mean field Hamiltonian can be diagonalized to solve for boson 
dispersions. 
For small boson density $\kappa$ the bosons will be gapped.
Increasing boson density will cause boson condensation at a critical boson
density $\kappa_c$, 
which corresponds to a magnetic ordering transition, 
and the details of the magnetic order can be derived from 
the structure of the boson condensates\cite{Sachdev}. 

For the Heisenberg model, the mean field ansatz can be solved from the 
self-consistent equations, 
\be
\bra \hat{A}_{ij}\ket_{\rm MF}=-A_{ij}/J_{ij},\quad
\bra \hat{B}_{ij}\ket_{\rm MF}=-B_{ij}/J_{ij},
\label{equ:SCF}
\ee
together with the average constraint \Equ{equ:averageconstraint}. 
Self-consistent equations for non-Heisenberg models 
can in principle be derived as well.  

As discussed in Ref.~\cite{SBPSG}, 
for the emergent gauge theory to be Z$_2$, 
it will need either both ansatz $A_{ij}$ and $B_{ij}$, 
or only ansatz $A_{ij}$ but with geometric frustration. 
Nearest-neighbor ansatz $A_{<ij>}$ on honeycomb lattice is bipartite and 
will lead to a U(1) gauge theory.  
Since the spin Hamiltonian \Equ{equ:Hspin} have strong 
further neighbor couplings, 
it is natural to assume that 
next-nearest-neighbor $A_{<<ij>>}$ is nonzero, 
which is sufficient to ``Higgs'' the U(1) gauge field into Z$_2$.

\section{Projective Symmetry Group of Schwinger Boson Mean Field Theories 
on Honeycomb Lattice}
\label{sec:SBPSG}
The mean field theory \Equ{equ:SBMF} is not invariant under
the local U(1) gauge transformations of the Schwinger bosons
\be
b_{j\alpha}\to e^{\im \phase(i)}b_{j\alpha},\ \alpha=\us,\ds
\ee
where the phase $\phase(j)$ can depend on site $j$. 
The ansatz will transform accordingly as  
\be
 A_{ij}\to e^{\im[\phase(i)+\phase(j)]}A_{ij},\quad
 B_{ij}\to e^{\im[-\phase(i)+\phase(j)]}B_{ij}
\ee
However the physical spin state is gauge invariant if 
the constraint \Equ{equ:constraint} is implemented exactly. 
Thus different mean field ansatz may correspond to the same physical state. 
Moreover the physical symmetries, e.g. the space group symmetry,  
may not be explicitly present in the mean field ansatz. 
And it is not straightforward to test whether a given mean field ansatz 
actually conforms all the physical symmetries under the 
constraint \Equ{equ:constraint}.
It was first noted by Wen and collaborators, in the studies of 
fermionic mean field theories of spin liquids, 
that the mean field theory should have a projective 
symmetry\cite{WenPSG,WenZhouPSG}. 
Namely the mean field ansatz should be invariant under a 
combined physical symmetry group and gauge group operation, 
a projective symmetry group operation. 
The structure of the physical symmetry group constrains possible 
structures of this projective symmetry group, 
thus constrains possible spin liquid states. 
This idea was generalized to Schwinger boson states in Ref.~\cite{SBPSG}
and applied to triangular and kagome lattices.
Here we will directly apply it to honeycomb lattice. 
More detailed discussion of the formalism can be found in Ref.~\cite{SBPSG}.

\begin{figure}
\includegraphics{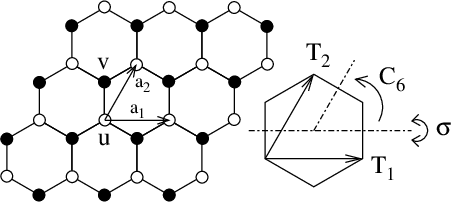}
\caption{The honeycomb lattice is shown on the left. 
Open(filled) circles indicate the two sublattices. 
$\veca_1,\veca_2$ are primitive vectors. 
For simplicity we assume the lattice constant $a=|\veca_1|=|\veca_2|=1$. 
$u,v$ denote the two sites within one unit cell. 
The hexagon on the right is the enlarged unit cell with 
schematic illustration of the space group generators,
translations $T_1$ and $T_2$, six-fold rotation $C_6$, 
and reflection $\mirror$.
}
\label{fig:lattice}
\end{figure}

The honeycomb lattice and its space group generators are 
illustrated in \Fig{fig:lattice}. 
Sites are labeled as $(x,y,w)$ with integer $x,y$ indicating 
the unit cell at $x\veca_1+y\veca_2$, 
and $w=u,v$ indicates the two sites in the unit cell. 

The space group of honeycomb lattice is generated by two translations
 $T_1$ along $\veca_1$, 
and $T_2$ along $\veca_2$, 
and a counter-clockwise six-fold rotation $C_6$ around 
the hexagon center $(1/3)(\veca_1+\veca_2)$, 
and a reflection $\mirror$ around the horizontal axis through 
the same hexagon center. Their actions on the lattice are
\begin{subequations}
\bea
T_1& : & (x,y,w) \to (x+1,y,w),\ w=u,v\\
T_2& : & (x,y,w) \to (x,y+1,w),\ w=u,v\\
C_6& : & \left \{
 \begin{array}{rcl}
(x,y,u) &\to& (-y+1,x+y-1,v)\\
(x,y,v) &\to& (-y,x+y,u)
 \end{array}\right. \\
\mirror & : & 
 \left \{
 \begin{array}{rcl}
(x,y,u) &\to& (x+y,-y,v)\\
(x,y,v) &\to& (x+y,-y,u)
 \end{array}\right. 
\eea
\end{subequations}

We associate a U(1) gauge group element,
$e^{\im \phase_X(j)}$ dependent on site $j$, 
to each element $X$ of the space group, 
and demand that 
the mean field ansatz be invariant under the combined PSG operation
\be
 b_{j\alpha}\to e^{\im \phase_X[X(j)]} b_{X(j)\alpha},\ \alpha=\us,\ds
\ee
where $X(j)$ is the image of site $j$ under the action of $X$. 
The structure of the space group can be used for solving the allowed 
phase functions $\phase_X(j)$. The solution is straightforward and 
listed in \App{app:PSG}. In the end we have 
\be
\begin{split}
\phase_{T_1}(x,y,w)\ & =  0, \quad
\phase_{T_2}(x,y,w)\  =  p_1\pi x, \\
\phase_{C_6}(x,y,w)\ & = 
 p_1 \pi \frac{x(x+2y-1)}{2}+\frac{(p_7+p_8+p_9)\pi}{2}, \\
\phase_{\mirror}(x,y,u)\ & = 
 p_1 \pi [\frac{y(y-1)}{2}+x]+p_1 \pi y+\frac{(p_7+p_9)\pi}{2}, \\
\phase_{\mirror}(x,y,v)\ & =
 p_1 \pi [\frac{y(y-1)}{2}+x]+p_1 \pi y+\frac{(p_7-p_9)\pi}{2}.
\end{split}
\label{equ:PSGSolution}
\ee
with $w=u,v$ labels sublattice, and four free integer parameters 
$p_1,p_7,p_8,p_9=0,1\mod 2$. 
Therefore there are at most 16 Z$_2$ states. 
Requiring nonzero nearest-neighbor $A_{<ij>}$, 
which is natural for strong nearest-neighbor Heisenberg AFM coupling, 
eliminates two parameters, $p_7=1$ and $p_9=p_8$.
If next-nearest-neighbor $A_{<<ij>>}$ is also nonzero 
as discussed in the end of \Sec{sec:SBMFT}, 
one more paremeter can be eliminated, $p_8=1$, 
and we are left with only one free parameter $p_1=0,1$. 
So there are only two relevant Z$_2$ states with  
 \be
 \begin{split}
 \phase_{T_1}(x,y,w)\ & =  0, \quad
 \phase_{T_2}(x,y,w)\  =  p_1\pi x, \\
 \phase_{C_6}(x,y,w)\ & = 
 p_1 \pi \frac{x(x+2y-1)}{2}-\frac{\pi}{2}, \\
 \phase_{\mirror}(x,y,u)\ & = 
  p_1 \pi [\frac{y(y-1)}{2}+x+y]+\pi, \\
 \phase_{\mirror}(x,y,v)\ & =
 p_1 \pi [\frac{y(y-1)}{2}+x+y]. 
 \end{split}
\label{equ:PSGSolution2}
\ee

\begin{figure}
\includegraphics{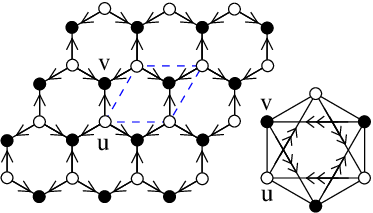}
\caption{(Color online) The zero-flux ansatz. 
Left part shows the nearest-neighbor ansatz. 
Single arrow from $i$ to $j$ means $A_{<ij>}=-A_{<ji>}=A_1>0$. 
All nearest-neighbor $B_{<ij>}$ must be zero according to \App{app:ansatz}. 
Blue dash rhombus is the unit cell of the mean field theory, 
containing two sites $u,v$. 
The large hexagon on the right is the enlarged mean field unit cell
showing the next-nearest-neighbor bonds. 
Double arrow from $i$ to $j$ means $A_{<<ij>>}=-A_{<<ji>>}=A_2$. 
All next-nearest-neighbor $B_{<<ij>>}=+B_2$ are real according 
to \Equ{equ:B2-1}-\Equ{equ:B2-6}.
}
\label{fig:zeroflux}
\end{figure}

\begin{figure}
\includegraphics{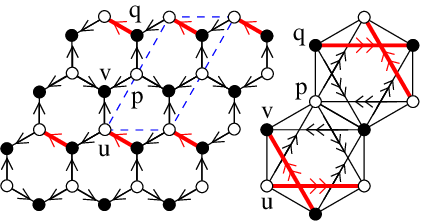}
\caption{(Color online) The $\pi$-flux ansatz. 
Left part shows the nearest-neighbor ansatz. 
Single arrow from $i$ to $j$ means $A_{<ij>}=-A_{<ji>}=A_1>0$.
All nearest-neighbor $B_{<ij>}$ must be zero according to \App{app:ansatz}.
Blue dash rhombus is the doubled unit cell of the mean field theory, 
containing four sites $u,v,p,q$. 
The large double hexagon on the right is the enlarged mean field unit cell
showing the next-nearest-neighbor bonds. 
Double arrow from $i$ to $j$ means $A_{<<ij>>}=-A_{<<ji>>}=A_2$.
All next-nearest-neighbor $B_{<<ij>>}=\pm B_2$ are real, 
with the $\pm$ signs given in \Equ{equ:B2-1}-\Equ{equ:B2-6}. 
Red thick bonds are those different from the zero-flux ansatz 
\Fig{fig:zeroflux}.
}
\label{fig:piflux}
\end{figure}

From the solutions of PSG one can construct all symmetry allowed 
mean field ansatz. 
The expressions of $A_{ij}$ up to fourth neighbors 
and $B_{ij}$ up to next-nearest-neighbor 
are listed in \App{app:ansatz}. 
The nearest-neighbor and next-nearest-neighbor $A_{ij}$ 
are also illustrated in \Fig{fig:zeroflux} and \Fig{fig:piflux}
for zero- and $\pi$-flux states respectively. 
In this paper the magnitudes of nearest-neighbor $|A_{<ij>}|$ and 
next-nearest-neighbor $|A_{<<ij>>}|$ are denoted as 
$A_1$, $A_2$ respectively. 
The two states are more intuitively distinguished by the gauge-invariant 
flux\cite{Tchernyshyov} in the elementary hexagon, defined as 
the phase of $A_{ij}(-A_{jk}^*)A_{k\ell}(-A_{\ell m}^*)A_{mn}(-A_{ni}^*)$, 
where the six sites $i,j,k,\ell,m,n$ are around a hexagon. 
For these two states 
this flux is $p_1 \pi$ so the time-reversal symmetry is 
also satisfied. 

\section{Z$_2$ Spin Liquids on Honeycomb Lattice}
\label{sec:fluxstates}
In this Section we study, within the mean field treatment, 
some simple properties of the two Z$_2$ spin liquid states found 
through the PSG analysis. 
For simplicity we will only use nearest-neighbor $A_{<ij>}=\pm A_1$ 
and next-nearest-neighbor bonds $A_{<<ij>>}=\pm A_2$, 
with $A_1$ real positive. 
The $\pm$ signs are given in \Fig{fig:zeroflux} and \Fig{fig:piflux}.
Because the spin Hamiltonian is very complicated, 
we will not compute energetics of these states 
and will not derive/solve self-consistent equations of ansatz $A_1,A_2$. 
Instead we will treat the ratio $A_2/A_1$ as a variational parameter 
and study the ``phase diagram'' with respect to it. 
This parameter can in principle be tuned by, for example, 
the $J_2/J_1$ ratio in the nearest-neighbor next-nearest-neighbor
$J_1$-$J_2$ Heisenberg AFM model on honeycomb lattice,
which is proportional to $(t/U)^2$ for small $t/U$ 
[see {e.g.} \Equ{equ:Hspin}]. 

Note that the $J_1$-$J_2$ Heisenberg model on honeycomb lattice has been 
studied within a Schwinger boson formalism by Mattsson {\etal}\cite{Mattsson}. 
However only the nearest-neighbor $A_{<ij>}$ and 
next-nearest-neighbor $B_{<<ij>>}$ were used. 
So that theory has U(1) gauge field instead of Z$_2$ and will be unstable. 
More recently Cabra {\etal}\cite{Cabra} studied a $J_1$-$J_2$-$J_3$ model 
with $J_3=J_2$ using Schwinger boson mean field theory. 
They found a commensurate colinear magnetic order with large $J_2/J_1$,
which is different from
 the incommensurate order obtained in 
the present paper with large $A_2/A_1$ in the zero-flux state,  
The small $J_2/J_1$ region of phase diagram in Ref.~\cite{Cabra} 
qualitatively agrees with our small $A_2/A_1$ region for 
the zero-flux state in \Fig{fig:kappac-zeroflux}.

\subsection{The Zero-flux State}
\label{ssec:zeroflux}
The zero-flux Z$_2$ spin liquid (\Fig{fig:zeroflux}) is a promising candidate 
for the numerically observed short-range RVB state. 
It has gapped bosonic spinons coupled to Z$_2$ gauge field. 
And it has a continuous transition into the N\'eel order 
even with small nonzero next-nearest-neighbor mean field coupling 
$A_2$, as long as $A_2 < A_1/2$. 
The continuum field theory close to this transition 
is derived following the method in Ref.~\cite{SachdevEffectiveTheory}. 
The effective theory shows a nontrivial coupling of bosons to 
the Higgs field involving cubic power of spatial derivatives, 
which allows a direct transition from Z$_2$ spin liquid to 
N\'eel order. 
This is in contrast to the conventional theory of transiton between Z$_2$ 
spin liquid and magnetic ordered state\cite{SachdevRead} 
which will generically give a non-colinear incommensurate magnetic order. 

The unit cell of \Fig{fig:zeroflux} contains two sites $u,v$. 
Fourier transform the bosons on each sublattice ($w=u,v$), 
\be
b_{(x,y,w)\alpha}=\frac{1}{\sqrt{ N_{\rm unit\ cells}}}
\sum_{\vk} e^{-\im (k_1 x+k_2 y)}b_{\vk w \alpha}
\ee
where $k_{1,2}\equiv \vk\cdot \veca_{1,2}$, 
the mean field Hamiltonian \Equ{equ:SBMF} becomes, up to a constant, 
\be
H_{\rm MF}=\sum_{\vk}
\Psi_{\vk}^\dagger
\begin{pmatrix}
\mu \mathbf{1}_{2\times 2} & A_1 P_1+A_2 P_2\\
-A_1 P_1-A_2^* P_2 & \mu \mathbf{1}_{2\times 2} 
\end{pmatrix}
\Psi_{\vk}^\nd
\label{equ:zeroflux-MF}
\ee
where $\Psi_{\vk}$ is a four component field 
$\Psi_{\vk}=(
 b_{\vk u \us}^\nd,b_{\vk v \us}^\nd,
 b_{-\vk, u \ds}^\dagger, b_{-\vk, v \ds}^\dagger)^T$ 
(superscript $^T$ means transpose), 
$\mathbf{1}_{2\times 2}$ is $2\times 2$ identity matrix, 
$P_{1,2}(\vk)$ are $2\times 2$ anti-hermitian matrices,
\be
P_1(\vk)=\begin{pmatrix}
 0 & \frac{+1+e^{\im(k_1-k_2)}+e^{-\im k_2}}{2}\\
\frac{-1-e^{\im(k_2-k_1)}-e^{\im k_2}}{2} & 0
\end{pmatrix}.
\ee
and 
\be
P_2(\vk)=\im [\sin(k_2)-\sin(k_1)+\sin(k_1-k_2)]\mathbf{1}_{2\times 2}.
\ee

\begin{figure}[t]
\includegraphics{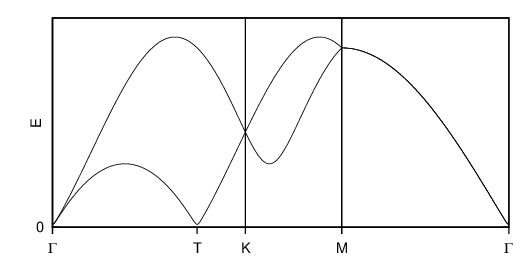}
\caption{
The zero-flux mean field boson dispersion $E_{\pm}$ 
\Equ{equ:dispersion-zeroflux}, 
with $A_2/A_1=1/2$ and average boson density $\bra\hat{n}\ket_{\rm MF}=1$ 
(for spin-1/2 model),
along high symmetry directions $\Gamma$-$K$-$M$-$\Gamma$
[see \Fig{fig:BZ}(a)]. Note the very low energy boson modes at $T$ point.
}
\label{fig:dispersion-zero}
\end{figure}

The mean field Hamiltonian can be diagonalized by a Bogoliubov 
transformation\cite{Sachdev}. 
The mean field dispersion has two branches $E_{\pm}$,  
each is doubly degenerate, 
\be
E_{\pm}(\vk)=\sqrt{\mu^2-A_1^2 f_1
\mp 2 A_1 \Re A_2 \sqrt{f_1}f_2
-|A_2|^2 (f_2)^2
}
\label{equ:dispersion-zeroflux}
\ee
where 
$f_1=[3+2\cos(k_1)+2\cos(k_2)+2\cos(k_1-k_2)]/4$, 
$f_2=4 \sin(k_1/2)\sin(k_2/2)\sin[(k_1-k_2)/2]$, 
$\Re A_2$ is the real part of $A_2$. 
An example of the dispersion is shown in \Fig{fig:dispersion-zero}.

When the dispersion is gapped, $E_{\pm}>0$, 
the average boson number $\kappa\equiv \bra\hat{n}\ket_{\rm MF}$ is
\be
\kappa=\int \frac{\dif k_1 \dif k_2}{4\pi^2} 
\frac{1}{2}\left (\frac{|\mu|}{E_{+}(\vk)}+\frac{|\mu|}{E_{-}(\vk)}\right )-1
\ee
Since we want the system to be stable against magnetic ordering, 
we want to maximize its capability of containing bosons. 
When $A_1$ and magnitude $|A_2|$ are fixed, 
the above boson density will be maximized if $A_2$ is real. 
Therefore $A_2$ will be assumed as real positive hereafter 
(real negative $A_2$ case is related to real positive case 
by a gauge transformation). 

\begin{figure}
\includegraphics{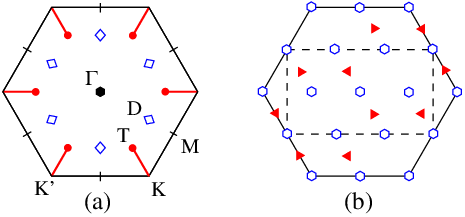}
\caption{(Color online) 
(a). Hexagon is the Brillouin zone of the zero-flux ansatz. 
Central black dot is the $\Gamma$ point $(k_1,k_2)=(0,0)$, 
where boson condensation happens when $A_2/A_1<1/2$. 
When $A_2/A_1$ increases from $1/2$ to $+\infty$, 
the boson condensation momenta move along the red short lines, 
$\pi < |\vk| < 4\pi/3$, from $T$ to $K$($K'$). 
The $\Gamma$ point, three BZ edge center $M$ points, 
six $T$ points(filled red circle, $|\vk|=\pi$) on $\Gamma$-$K$($K'$) lines, 
and six $D$ points(open blue diamond, $|\vk|=\sqrt{3}\pi/2$) on
 $\Gamma$-$M$ lines 
are the would-be magnetic Bragg peaks for zero-flux spin liquid 
with $A_2/A_1\sim 1/2$, 
namely peaks in dynamical spin susceptibility at low frequency around spin gap.
(b). Hexagon is the Brillouin zone of the original lattice. 
Dash rectangle is the reduced Brillouin zone for the $\pi$-flux ansatz. 
In the $\pi$-flux state, bosons can condense at the momenta indicated by the 
filled red small triangles, and produce magnetic Bragg peaks with possible 
wave vectors indicated by the open blue small hexagons. 
}
\label{fig:BZ}
\end{figure}

When $A_2<A_1/2$ 
the dispersion minimum is at the $\Gamma$ point, $(k_1,k_2)=(0,0)$, 
in the Brillouin zone(BZ) [see \Fig{fig:BZ}(a)]. 
With increasing boson density the bosons will finally condense at 
the $\Gamma$ point. 
Like in the triangular and kagome case\cite{Sachdev,SBPSG}, 
the structure of the condensate can be determined
by solving the eigenvectors of \Equ{equ:zeroflux-MF} with zero eigenvalues 
at the condensation momenta. 
Let $(k_1,k_2)=(0,0)$ in \Equ{equ:zeroflux-MF} and demand (one of) $E_{\pm}$ 
to be zero, we get $|\mu/A_1|=3/2$ and two eigenvectors
$(1,0,0,-1)^T$, $(0,1,1,0)^T$ corresponding to the zero eigenvalues. 
Therefore the condensate at this momentum is a linear combination of these 
two vectors,
\be
\bra \Psi_{\vk=(0,0)} \ket
=z_1 (1,0,0,-1)^T+z_2 (0,1,1,0)^T
\ee
Complex numbers $z_1$, $z_2$ determine the orientation of staggered moments,
as in the case of triangular lattice\cite{Sachdev}. 
Define $z=(z_1,z_2^*)^T$, then the Schwinger bosons on 
sublattice $u$($v$) becomes $\bra b_{\alpha}\ket=z$
($\bra b_{\alpha} \ket=\im \sigma^y z^*$).
The moment on sublattice $u$($v$) is 
$\vm_u=(1/2)z^\dagger \vsigma^\nd z^\nd$ 
[$ \vm_v=(1/2)z^T (-\im \sigma^y)\vsigma (\im \sigma^y) z^*=
-(1/2)z^\dagger \vsigma^\nd z^\nd = -\vm_u$]. 
This is the N\'eel order. 

At $A_2/A_1=1/2$, the minima of dispersion jump to
six $T$ points on the $\Gamma$-$K$($K'$) lines with $|\vk|=\pi$ 
[BZ corner $K$($K'$) point has $|\vk|=4\pi/3$]. 
Further increase $A_2/A_1$ to $+\infty$ will move the minima toward   
the $K$($K'$) points [see \Fig{fig:BZ}(a)]. 
The boson condensation in this case will in general lead to 
incommensurate magnetic order.
Note that the $A_2/A_1=+\infty$ limit is just two copies of 
decoupled zero-flux triangular lattice Schwinger boson mean field 
theory\cite{Sachdev,SBPSG}.

A mean field ``phase diagram'' in terms of the variational parameter $A_2/A_1$
and average boson density is constructed as \Fig{fig:kappac-zeroflux}. 
There is a very small parameter range $0.493 < A_2/A_1 < 0.516$
where the critical boson density is greater than unity,
namely the spin-1/2 system will remain to be a gapped spin liquid. 
This is particularly promising for explaining the numerically observed 
transition from short-range RVB to N\'eel state as $U/t$ is increased. 
Because increasing of $U/t$ will decrease $J_2/J_1\propto (t/U)^2$, 
and thus decrease $A_2/A_1$,  
the spin-1/2 system will move to the left along 
the dash line in \Fig{fig:kappac-zeroflux},  
and cross the mean field phase boundary between 
the zero-flux Z$_2$ spin liquid and N\'eel order. 

In this scenario, the spin liquid will be very close to 
the mean field tricritical point 
$A_2/A_1=1/2$ and $\bra\hat{n}\ket_{\rm MF}\approx 1.18$. 
Therefore the momenta of low energy bosons are not only the 
$\Gamma$ point, but also the six $T$ ($|\vk|=\pi$) points in \Fig{fig:BZ}(a). 
The dispersion for $A_2/A_1=1/2$ and $\bra\hat{n}\ket_{\rm MF}=1$ (spin-1/2) 
case is drawn along high symmetry directions in \Fig{fig:dispersion-zero} 
to illustrate this point. 
The dynamical spin susceptibility at low frequency around the spin gap
will have peaks at wave vectors connecting two(can be the same) 
boson condensation momenta, 
these include not only the $\Gamma$ point, 
but also three Brillouin zone edge center $M$ points, 
and these six $T$ points, 
and six other $D$ points [\Fig{fig:BZ}(a)].

\subsection{Critical Field Theory for the Transition from Zero-flux State
to N\'eel Order}
Considering the spatial-temporal fluctuations of 
the would-be boson condensate $z$ in the zero-flux state 
close to the transition into N\'eel order, 
one can derive the critical field theory. 
The detailed derivation is given in \App{app:fieldtheory}. 
The boson part of the Lagrangian reads
\be
\begin{split}
& \mathcal{L}_{z} =
\int \dif^2 \vecr \Big \{ 
|\deriv_\tau z|^2+c^2|\deriv_\vecr z|^2+m^2 |z|^2
\\ & \quad
+\lambda_3\, z^*[\sum_{j=1}^{3}(\vece_j\cdot \deriv_{\vecr})^3]z+{c.c.}
\\ &\quad
+\lambda_H\, \Phi\cdot z^T (\im\sigma^y) [ 
\sum_{j=1}^{3}
(\vecd_j\cdot \deriv_\vecr)^3
]
z+{c.c.}
\Big \}
\end{split}
\label{equ:action}
\ee
where $\tau$ is the imaginary time, $\vecr$ is the spatial coodinates, 
$\Phi\sim A_2$ is the scalar Higgs field, 
${c.c.}$ means complex conjugate of the previous term, 
and $\deriv$ is 
the covariant derivative with minimal coupling 
to the compact U(1) gauge field coming from the 
Schwinger boson representation. 
Vectors
$\vece_1=(2\veca_2-\veca_1)/3$,  
$\vece_2=-(\veca_2+\veca_1)/3$,
$\vece_3=(2\veca_1-\veca_2)/3$, 
$\vecd_1=-\veca_1$, $\vecd_2=\veca_2$, 
$\vecd_3=\veca_1-\veca_2$ are defined for convenience. 
The velocity $c$ and boson mass $m$ and coupling constants
$\lambda_3$ and $\lambda_H$
can in principle be derived from 
the microscopic theory. 
Magnetic ordering transition happens when the mass $m$ vanishes. 

The transformation rules of  $z$ and $\Phi$ fields under space group symmetry 
can be derived from the zero-flux ($p_1=0$) PSG \Equ{equ:PSGSolution2}, 
\begin{subequations}
\bea
T_1,\ T_2 & : & z\to z,\quad \Phi\to\Phi, \\
C_6& : & z\to -\sigma^y z^*,\quad \Phi\to \Phi,\\
\mirror & : & z\to -\im \sigma^y z^*,\quad \Phi\to \Phi.
\eea
\end{subequations}
The Higgs field $\Phi\sim A_2$ transforms trivially. 
The Lagrangian \Equ{equ:action} is invariant under the PSG. 

Note that the form of the coupling between bosons $z$ and 
the Higgs field $\Phi$ is constrained by the PSG,
namely the microscopic lattice symmetry. 
It is very different from the typical coupling\cite{SachdevRead} 
which involves only one spatial derivative, 
such coupling would violate the six-fold rotation symmetry here. 
Naive power counting shows that this coupling here, 
with cubic power of spatial derivatives, is irrelevant, 
which means the Higgs field will dynamically decouple from 
the bosons at low energy. 
Considering the anomalous dimensions will not change
this conclusion. 
This is why the Z$_2$ state here still 
produces a commensurate N\'eel order upon boson condensation
in contrast to the conventional theory\cite{SachdevRead}
where it usually becomes a non-colinear incommensurate order. 
However the Higgs mechanism for reducing U(1) to Z$_2$ is still intact, 
as long as the Higgs condensate $\Phi\sim A_2$ is nonzero, 
providing stability against confinement in compact U(1) gauge theory in 
$2+1$ dimension. 
It would be very interesting to see if the same critical field theory 
can be reached from the N\'eel order side. 

At the transition point, the low energy theory is 
the O(4) invariant critical theory for the 
transition between a spiral magnet and a 
gapped spin liquid\cite{ChubukovPRL,ChubukovNPB,Isakov}. 
The scaling properties have been studied within 
large-$N$ expansion\cite{ChubukovPRL,ChubukovNPB} and 
also numerically\cite{Isakov}. 
For example spin-spin correlations will 
have power-law scaling at large distance
\be
\bra \vecS(0)\cdot \vecS(\vecr)\ket \sim |\vecr|^{-\eta}
\ee
where $\eta$ has been numerically determined\cite{Isakov} as $\eta=1.373(3)$. 
This can be checked with the finite-size scaling results of the Hubbard model
when $U/t$ is tuned to the magnetic ordering transition. 

\newcommand\p{$\pi$}
\subsection{The {\p}-flux State}
\label{ssec:piflux}
Now we consider the $\pi$-flux state in \Fig{fig:piflux}. 
The unit cell for the mean field theory is doubled along $\veca_2$ 
direction and contains four sites $u,v,p,q$. 
The Brillouin zone is halved as shown in \Fig{fig:BZ}(b). 
However we stress here that the physical spin state obtained from 
imposing the constraint \Equ{equ:constraint} on this mean field wave function 
has the original translation symmetry of honeycomb lattice, 
and this is guaranteed by the PSG. 

The mean field Hamiltonian after Fourier transform looks like, 
up to a constant, 
\be
\sum_{\vk} \Psi_{\vk}^\dagger
\begin{pmatrix}
\mu \mathbf{1}_{4\times 4} & A_1 P_1+A_2 P_2\\
-A_1 P_1 - A_2^* P_2 & \mu \mathbf{1}_{4\times 4}
\end{pmatrix}
\Psi_{\vk}^\nd
\label{equ:piflux-MF}
\ee
where $\Psi_{\vk}$ is an eight component field, 
$\Psi_{\vk}=(
 b_{\vk u \us}^\nd,b_{\vk v \us}^\nd,b_{\vk p \us}^\nd,b_{\vk q \us}^\nd,
 b_{-\vk, u \ds}^\dagger, b_{-\vk, v \ds}^\dagger,
b_{-\vk, p \ds}^\dagger, b_{-\vk, q \ds}^\dagger)^T$, 
$\mathbf{1}_{4\times 4}$ is $4\times 4$ identity matrix, 
$P_{1,2}$ are $4\times 4$ anti-hermitian matrices,
\be
P_1=\frac{1}{2}\begin{pmatrix}
0 & 1 & 0 & -\epsilon_3^{-1} + \epsilon_2\\
-1 & 0 & -1-\epsilon_1^{-1} & 0\\
0 & 1+\epsilon_1 & 0 & 1\\
\epsilon_3 - \epsilon_2^{-1} & 0 & -1 & 0
\end{pmatrix}.
\ee
\begin{widetext}
\be
P_2=\frac{1}{2}\begin{pmatrix}
2\im\sin(k_1) & 0 & 1-\epsilon_2-\epsilon_1^{-1}-\epsilon_3^{-1} & 0\\
0 & -2\im\sin(k_1) & 0 & 1-\epsilon_2-\epsilon_1^{-1}-\epsilon_3^{-1}\\
-1+\epsilon_2^{-1}+\epsilon_1+\epsilon_3 & 0 & -2\im\sin(k_1) & 0\\
0 & -1+\epsilon_2^{-1}+\epsilon_1+\epsilon_3 & 0 & 2\im\sin(k_1)
\end{pmatrix}.
\ee
\end{widetext}
with the short-hand notations 
$\epsilon_1=e^{\im k_1}$, $\epsilon_2=e^{-\im k'_2}$, 
$\epsilon_3=e^{\im (k'_2 - k_1)}$, 
and $k_1\equiv \vk\cdot \veca_1$, $k'_2\equiv \vk\cdot(2\veca_2)$. 
Note that $k'_2$ is twice of the $k_2$ in previous Subsection. 

The mean field Hamiltonian can in principle be diagonalized by 
a Bogoliubov transformation to give the mean field dispersion. 
However with $A_1$ and $A_2$ both nonzero  
this is very difficult analytically. 
In the following we will set $A_2$ to zero 
and present some results for the nearest-neighbor ansatz. 
The mean field dispersion with only nearest-neighbor ansatz 
has two branches, each is four-fold degenerate, 
\be
E_{\pm}^{(\pi)}(\vk)=\sqrt{\mu^2-
A_1^2
[
3/4\pm
\sqrt{ f(\vk)}
]
}
\ee
where $f(\vk)=[3+\cos(2k_1)+\cos(k'_2)-\cos(2k_1-k'_2)]/8$. 

Average boson density $\kappa\equiv \bra \hat{n}\ket_{\rm MF}$ is 
\be
\kappa=\int \frac{\dif k_1 \dif k'_2}{4\pi^2}
 \frac{1}{2}\left (\frac{|\mu|}{E_+^{(\pi)}(\vk)}
+\frac{|\mu|}{E_-^{(\pi)}(\vk)}\right)-1
\ee
The critical boson density is achieved when $|\mu/A_1|=\sqrt{3/2}$,
and $\kappa_c=2.14 > 1$. 
Taken at face value it means this state can remain quantum disordered 
for spin-1/2 and even spin-1 systems. 

The bosons will condense at four momenta in the reduced Brillouin zone
[see \Fig{fig:BZ}(b)], which are 
$\vk=\pm\vk_{c1}=\pm(k_1=\pi/6,k'_2=-\pi/3)$
and $\vk=\pm\vk_{c2}=\pm(k_1=-5\pi/6,k'_2=-\pi/3)$.
The condensate at each momentum will be
\begin{subequations}
\bea
\bra \Psi_{\vk=+(\pi/6,-\pi/3)} \ket
& = & z_1 V_1+z_2 V_2, \\
\bra \Psi_{\vk=-(\pi/6,-\pi/3)} \ket
& = & w_1 V_1^*+w_2 V_2^*, \\
\bra \Psi_{\vk=+(-5\pi/6,-\pi/3)} \ket
& = & z_3 V_3+z_3 V_4, \\
\bra \Psi_{\vk=-(-5\pi/6,-\pi/3)} \ket
& = & w_3 V_3^*+w_4 V_4^* .
\eea
\end{subequations}
with complex coefficents $z_{1,2,3,4},w_{1,2,3,4}$, 
and the complex vectors $V_1,V_2$ are eigenvectors of 
\Equ{equ:piflux-MF} at $\vk_{c1}=(\pi/6,-\pi/3)$ with eigenvalue zero, 
and $V_3,V_4$ are for $\vk_{c2}=(-5\pi/6,-\pi/3)$. 
The vectors $V_{1,2,3,4}$ are explicitly given below,
\be
\begin{split}
V_1 &= (e^{-\im \pi/12},0,\sqrt{2+\sqrt{3}},0,0,-e^{-\im \pi/12}\sqrt{2+\sqrt{3}},0,-1),\\
V_2 &= (0,e^{-\im\pi/12}\sqrt{2+\sqrt{3}},0,-1,e^{-\im \pi/12},0,\sqrt{2+\sqrt{3}},0),\\
V_3 &= (e^{5\im \pi/12}\sqrt{2+\sqrt{3}},0,1,0,0,-e^{5\im\pi/12},0,-\sqrt{2+\sqrt{3}}),\\
V_4 &= (0,e^{5\im\pi/12},0,\sqrt{2+\sqrt{3}},e^{5\im\pi/12}\sqrt{2+\sqrt{3}},0,1,0)
\end{split}
\ee
Note that $z_{1,2,3,4},w_{1,2,3,4}$ may not be independent, 
because one need to make sure that the number of condensed bosons 
on every site is the same\cite{Sachdev}. 

The magnetic order is complicated but will certainly not be the Neel order. 
Because bosons have to condense at several different momenta
otherwise the condensed boson density(size of the magnetic moment) would be 
non-uniform on the four sublattices. 
Without knowing the detailed condensate structure we can still determine 
the possible magnetic Bragg peak wavevectors, which are the differences 
between two boson condensation momenta. 
These possible magnetic Bragg peaks are
$(k_1,k_2)=\pm(\pi/3+m\pi,-\pi/3+n\pi),\pm(m\pi,n\pi)$ with integers $m,n$ 
and are illustrated in \Fig{fig:BZ}(b). 
These momenta are accessible on $6\times 6$, $12\times 12$ and $18\times 18$ 
lattices used in the quantum Monte Carlo study\cite{QMC}. 
So whether this $\pi$-flux state is realized can be tested 
by measuring static spin structure factor at these momenta 
in the magnetic ordered phase. 
The detailed magnetic order pattern will be very nontrivial 
like that from the triangular lattice $\pi$-flux state\cite{SBPSG}, 
but will be left for future works. 

We will not study the effect of the next-nearest-neighbor coupling $A_2$
in the $\pi$-flux state in this paper. 
We just note here that with $A_2/A_1\to\infty$, the mean field ansatz
\Fig{fig:piflux} becomes two copies of decoupled 
$\pi$-flux states on the triangular lattice found in Ref.~\cite{SBPSG}. 

It would be interesting to realize this $\pi$-flux state in 
a simple spin model on honeycomb lattice. 
However for the nearest-neighbor Heisenberg model
general argument\cite{Tchernyshyov} indicates that 
zero-flux state will always have lower energy than the $\pi$-flux state. 
Ring-exchange interaction (for the six sites around a hexagon) 
may favor the $\pi$-flux state\cite{SBPSG}. 
However the natural sign of the ring-exchange coupling 
derived from the Hubbard model will actually favor the zero-flux state 
as discussed in Ref.~\cite{SBPSG}. 
Thus the $\pi$-flux state is not likely realized in 
the numerical simulation of the Hubbard model\cite{QMC}. 

\section{Conclusions}
\label{sec:conclusion}
In hope of understanding the numerical evidence of a short-range RVB state 
found by recent quantum Monte Carlo simulations of 
 honeycomb lattice Hubbard model\cite{QMC}, 
and the possibly continuous quantum phase transition from the short-range RVB 
to the magnetic ordered N\'eel state,
we studied the Z$_2$ spin liquids within the Schwinger boson mean field theory. 
Applying the projective symmetry group method for Schwinger boson 
states\cite{SBPSG} we completely classified possible 
Z$_2$ Schwinger boson spin liquid states on honeycomb lattice. 
Symmetry allowed mean field ansatz are derived 
for up to fourth neighbor couplings, 
which can be used for future studies of the Schwinger boson mean field theory. 
Assuming nonzero nearest-neighbor and next-nearest-neighbor
mean field couplings $A_1$ and $A_2$, 
there are only two Z$_2$ states on honeycomb lattice which 
do not break any lattice symmetry. 
The two states are differentiated by the gauge invariant flux, zero or $\pi$, 
in the elementary hexagon. 

The zero-flux state is a very promising candidate for the numerically 
observed short-range RVB state. 
Its critical boson density decreases from $1.18$ at $A_2/A_1=1/2$ 
to $0.516$ at $A_2/A_1=0$, and a continuous quantum phase transition 
to N\'eel order will happen in this process, 
emulating the behavior of the numerically studied Hubbard model 
when $U/t$ increase from below $U/t=4.3$ to $+\infty$. 
The critical field theory 
for the phase transition to N\'eel order 
is an O(4) invariant theory \Equ{equ:action}, 
with an irrelevant coupling between Higgs field and boson fields 
involving cubic power of spatial derivatives, 
unlike the conventional form of such coupling with only one spatial 
derivative\cite{SachdevRead}. 
Therefore it allows for a direct transiton 
from a Z$_2$ gapped spin liquid to a N\'eel order. 
In this scenario the spin liquid could have soft spin fluctuations 
at not only the ordering wave vector $\Gamma$ point, 
but also at Brillouin zone edge center $M$ points, 
and six $T$ ($|\vk|=\pi$) points, 
and six other $D$ points [see \Fig{fig:BZ}(a)].
which can be checked by numerically calculating the 
dynamical spin susceptibility. 
Also the magnetic ordering transition will be an O(4) invariant 
theory, the (finite-size) scaling of correlation functions 
can be checked against known results\cite{ChubukovPRL,ChubukovNPB,Isakov},
{e.g.} spin-spin correlation function behaves as $|\vecr|^{-1.373}$ 
at large distance $\vecr$.

The $\pi$-flux state has the critical mean field boson density
 $\kappa_c\approx 2.13$
(with only nearest-neighbor mean field couplings) well above unity. 
Boson condensation in the $\pi$-flux state will lead to magnetic Bragg peak 
at several wave vectors as show in \Fig{fig:BZ}(b), 
including the N\'eel order wave vector, 
which can be checked in the numerical simulations of 
the magnetic ordered phase. 
But for energetic reasons it is not likely realized in the Hubbard model. 

There are still many remaining interesting questions and 
possible future directions in this problem. 
(1). The Z$_2$ spin liquid on a torus will have four-fold 
ground state degeneracy which was not observed in 
the numerical simulation\cite{QMC}. 
It is possible that ground states in different topological sector 
actually carry different physical quantum number, e.g. quantum number 
with respect to six-fold rotation, 
thus not all of them were accessed in the simulation. 
It would be useful to work out these vison quantum numbers 
which can guide the search of topological order in the numerical work. 
(2). The critical field theory \Equ{equ:action} 
is derived from the spin liquid side. It would be very interesting  
to start from the N\'eel ordered side and see if the same conclusion 
can be reached. For comparison to numerics it may also be useful to compute 
the scaling properties of other observables. 
Also the mean field tricritical point in \Fig{fig:kappac-zeroflux}, 
where bosons condense at $\Gamma$ and six $T$ points, 
might also be of some interest. 
(3). The continuous Mott transition is not easy to understand with 
the Schwinger boson formalism, but is more natural 
in the fermionic spinon formulation. It may be interesting to study the 
Z$_2$ states with fermionic spinons, and see if a unified picture 
of both continuous Mott transition and magnetic ordering transition 
can be achieved. 
(4). It may be useful to derive the effective spin model 
from the Hubbard model to high orders of $t/U$, then 
compute energetics of the zero-flux Z$_2$ spin liquid state 
and other possible states, 
in order to produce a physical (mean field) phase diagram.   
(5). It may also be useful to have a concrete simple spin model 
which shows one of these Z$_2$ spin liquid ground states. 
$J_1-J_2$ model may be a good example, 
but unfortunately has sign problem preventing 
large scale quantum Monte Carlo simulations.  

There has been a proposal of non-magnetic insulator state in honeycomb 
Hubbard model close to the metal-insulator transition\cite{KouSP}. 
Its relation to the present study is however unclear yet.
Also in a recent paper by Xu and Sachdev\cite{Cenke} 
another Z$_2$ spin liquid state was proposed through a different formalism.  
Its relation to the Z$_2$ spin liquid studied here remains to be clarified.

\begin{acknowledgments}
The author thanks Ying Ran for bringing Ref.~\cite{QMC} to his attention, 
and acknowledges very helpful discussions with Ying Ran, Todadri Senthil, 
Ashvin Vishwanath, Xiao-Gang Wen, and Cenke Xu.
The author is especially grateful to Todadri Senthil for correcting 
the interpretation of the continuum field theory \Equ{equ:action}. 
The author is supported by the MIT Pappalardo Fellowship in Physics.
\end{acknowledgments}

\appendix

\section{Algebraic Solution of the Z$_2$ PSG on Honeycomb Lattice}
\label{app:PSG}
In this Appendix we list the detailed steps for solving 
the Z$_2$ PSGs on honeycomb lattice. 
The algebraic solutions will determine all possible symmetric Z$_2$ states
within the Schwinger boson formalism. 

The lattice and its space group generators are described in \Sec{sec:SBPSG}
and illustrated in \Fig{fig:lattice}. 
All independent commutation relations between the space group generators are
\be
\begin{split}
& 
T_1^{-1} T_2 T_1 T_2^{-1}=
T_1^{-1} T_2^{-1} T_1 T_2=
\\ &
T_1^{-1} C_6 T_1 T_2^{-1} C_6^{-1}=
T_2^{-1} C_6 T_1 C_6^{-1}=
C_6^6=
\\ &
T_1^{-1} \mirror T_1 \mirror^{-1}=
T_2^{-1} \mirror T_1 T_2^{-1} \mirror^{-1}=
\mirror^2=
\mirror C_6 \mirror C_6=
\SGid.
\end{split}
\ee
where $\SGid$ is the identity element of the space group. 

For reasons discussed in \Sec{sec:SBMFT} we will assume 
the invariant gauge group is Z$_2$. The generator of IGG is 
\be
\hat{b}_{j\alpha}\to -\hat{b}_{j\alpha},\ \alpha=\us,\ds,\ \forall{\rm site\ }j
\label{equ:IGGGenerator}
\ee

For each space group element $X$, associate a gauge group element [U(1) phase]
 $\exp[\im \phase_X(j)]$ such that the mean field Hamiltonian is 
invariant under the combined PSG operation
\be
b_{j\alpha}\to \exp[\im \phase_X(j)] b_{X(j)\alpha}
\ee
Note that these phases $\phase_X(j)$ and later equations of these phases
should be understood with implicit modulo $2\pi$. 

If a gauge transformation $b_{is}\to e^{\im \phase(i)} b_{is}$ is applied, 
then PSG elements transform as \cite{SBPSG}
 $\phase_X(i)\to \phase_X(i)+\phase(i)-\phase[X^{-1}(i)]$.
Using this gauge freedom one can always assume (on open boundary condition)
\be
\phase_{T_1}(x,y,w)=0,\quad \phase_{T_2}(x=0,y,w)=0
\label{equ:gaugefixing}
\ee
where $w=u,v$ labels sublattice, $(x,y)$ labels unit cell.

For simplicity of notations we define two forward finite differences
$\Delta_1 f(x,y)\equiv f(x+1,y)-f(x,y)$, and 
$\Delta_2 f(x,y)\equiv f(x,y+1)-f(x,y)$. 

From $T_1^{-1} T_2 T_1 T_2^{-1}=\SGid$, 
convert each space group element to its corresponding PSG element, 
the identity $\SGid$ to an unknown IGG element
$b_{i\alpha}\to e^{\im p_1\pi}b_{i\alpha}$, 
we have 
\be
\Delta_1 \phase_{T_2}(x,y,w)=p_1\pi
\ee
with integer $p_1=0,1\mod 2$.
Later used integers $p_{2,3,4,5,6,7,8,9}$ are also Z$_2$ integers. 
And equations between them should be understood with implicit modulo 2. 
Solution of this equation together with \Equ{equ:gaugefixing} is
\be
\phase_{T_2}(x,y,w)=p_1\pi x
\ee
From this one can already conclude that the flux in 
the elementary hexagon is $p_1\pi$. 

At this stage there are four remaining gauge freedoms. 
These gauge transformations do not change 
$\phase_{T_1}$, $\phase_{T_2}$ up to
IGG elements, but can be used to simplify other PSG elements.

Gauge freedom I: a global phase rotation, does not change any PSG elements,
\be
b_{(x,y,w)\alpha}\to e^{\im \phase}b_{(x,y,w)\alpha}
\ee
This can be used to fix one of the $A_{ij}$ to be real positive. 
We will fix $A_{(0,0,u)\to(0,0,v)}$ to be real positive.

Gauge freedom II: 
\be
b_{(x,y,w)\alpha}\to e^{\im \pi x}b_{(x,y,w)\alpha}
\ee

Gauge freedom III: 
\be
b_{(x,y,w)\alpha}\to e^{\im \pi (x+y)}b_{(x,y,w)\alpha}
\ee

Gauge freedom IV: staggered phase rotation,
\be
b_{(x,y,u)}\to e^{+\im \phase}b_{(x,y,u)},\quad
b_{(x,y,v)}\to e^{-\im \phase}b_{(x,y,v)},
\ee

From $T_1^{-1} C_6 T_1 T_2^{-1} C_6^{-1}=
T_2^{-1} C_6 T_1 C_6^{-1}=\SGid$ we have
\begin{subequations}
\bea
\Delta_1 \phase_{C_6}(x,y,w) & = & p_1\pi (x+y)+p_2\pi, \\
\Delta_2 \phase_{C_6}(x,y,w) & = & p_1\pi x +p_3\pi.
\eea
\end{subequations}
Its solution is
\be
\begin{split}
& 
\phase_{C_6}(x,y,w)
\\ 
= \ 
&
\phase_{C_6}(0,0,w)
 +p_1 \pi \frac{x(x+2y-1)}{2}
+p_2 \pi x+p_3 \pi y
\end{split}
\ee
If gauge freedom II is applied, $p_3$ becomes $p_3+1$,
therefore $p_3$ can always be assumed as zero.
If gauge freedom III is applied, $p_2$ becomes $p_2+1$, 
and $\phase_{C_6}(0,0,v)$ becomes $\phase_{C_6}(0,0,v)+\pi$, 
therefore $p_2$ can always be assumed as zero as well. 
If gauge freedom IV is applied, 
$\phase_{C_6}(0,0,u)$ becomes $\phase_{C_6}(0,0,u)+\phase$ 
and $\phase_{C_6}(0,0,v)$ becomes $\phase_{C_6}(0,0,v)-\phase$,
therefore $\phase_{C_6}(0,0,u)$ and $\phase_{C_6}(0,0,v)$ 
can always be assumed as the same. And now we have exhausted 
all gauge freedoms. 

From $T_1^{-1} \mirror T_1 \mirror^{-1}=
T_2^{-1} \mirror T_1 T_2^{-1} \mirror^{-1}=\SGid$ we have
\begin{subequations}
\bea
\Delta_1 \phase_{\mirror}(x,y,w) & = & p_4\pi, \\
\Delta_2 \phase_{\mirror}(x,y,w) & = & p_1 \pi y +p_5\pi.
\eea
\end{subequations}
Its solution is
\be
\phase_{\mirror}(x,y,w)=\phase_{\mirror}(0,0,w)
 +p_1 \pi y(y-1)/2+p_4 \pi x+p_5 \pi y
\ee

From $C_6^6=\SGid$ we have a constraint on $\phase_{C_6}(0,0,w)$, 
\be
3[\phase_{C_6}(0,0,u)+\phase_{C_6}(0,0,v)]+(p_1+p_2)\pi=p_6\pi
\ee

From $\mirror^2=\SGid$ we have a constraint on $\phase_{\mirror}(0,0,w)$,
\be
\phase_{\mirror}(0,0,u)+\phase_{\mirror}(0,0,v)+\pi(p_1 y^2+p_4 y)=p_7\pi
\ee
This ensures $p_4=p_1\mod 2$ because this equation is true for all $y$. 

From $\mirror C_6 \mirror C_6=\SGid$ we have a constraint on 
$\phase_{C_6}(0,0,w)$ and $\phase_{\mirror}(0,0,w)$, 
\be
\begin{split}
&
2\phase_{\mirror}(0,0,v)
+2\phase_{C_6}(0,0,u)
+p_1\pi (x+y)^2-p_5\pi (x+y)
\\
=\
&
2\phase_{\mirror}(0,0,u)
+2\phase_{C_6}(0,0,v)
+p_1\pi (x+y)^2-p_5\pi (x+y)
\\
& +(p_5-p_1)\pi
\\
=\
&
p_8\pi
\end{split}
\ee
This ensures $p_5=p_1$.
And we have
\be
\phase_{\mirror}(0,0,u)-\phase_{\mirror}(0,0,v)=p_9\pi
\ee
And the solution of $\phase_{C_6}(0,0,w)$ and $\phase_{\mirror}(0,0,w)$ is
\bea
\phase_{\mirror}(0,0,u) & = & (p_7+p_9)\pi/2\mod 2\pi, \\
\phase_{\mirror}(0,0,v) & = & (p_7-p_9)\pi/2\mod 2\pi, \\
\phase_{C_6}(0,0,w) & = & (p_7+p_8+p_9)\pi/2\mod 2\pi,
\eea
and $p_1+p_6+p_7+p_8+p_9=0\mod 2$ thus $p_6$ can be eliminated.

Considering all these constraints, 
$p_2=p_3=0$, $p_4=p_1$, and $p_6=p_1+p_7+p_8+p_9$,  
we will reach the final solution of PSG shown in the main text 
\Equ{equ:PSGSolution} 
with only five free Z$_2$ integer parameters $p_1,p_5,p_7,p_8,p_9$. 

\section{Realizations of the Z$_2$ PSG on Honeycomb Lattice: Mean Field Ansatz}
\label{app:ansatz}
In this Appendix we will use the solution of PSG to construct symmetry allowed
mean field ansatz. 
We will list the PSG allowed ansatz up to fourth neighbors
of the honeycomb lattice.

The algebraic solution of PSG is very general and usually contains many 
free parameters. 
When realized by a particular kind of ansatz, e.g. nearest-neighbor ansatz, 
the number of free parameter will be greatly reduced because there will be 
further constraints on the PSG.
For example, if $A_{ij}$ is nonzero, 
and there is a non-identity space group element $X$ such that
$X(i)=j,\ X(j)=i$, namely the bond $ij$ maps to its inverse $ji$, 
then $A_{ji}=-A_{ij}=\exp[\im\phase_X(j)+\im\phase_X(i)] A_{ij}$,
therefore $\phase_X(j)+\phase_X(i)=\pi\mod 2\pi$. 
All such independent non-identity space group elements $X$, 
which map $ij$ to itself or its inverse, need to be checked. 
The ansatz $A_{ij}$ is compatible with this PSG 
if all such checks are passed. Then ansatz on all symmetry related bonds
can be generated by applying the PSG operations. 

Nearest-neighbor ansatz $A_{<ij>}$: 
Assume $A_{(0,0,u)\to(0,0,v)} = A_1 > 0$. 
This bond under $\mirror$ becomes its inverse $(0,0,v)\to(0,0,u)$, 
then $\phase_{\mirror}(0,0,u)+\phase_{\mirror}(0,0,v)=\pi$, 
therefore $p_7=1$.
This bond under $T_1^{-1} C_6^3$ becomes its inverse as well, 
then 
$\phase_{C_6}(0,0,u)+2\phase_{C_6}(1,-1,v)
+2\phase_{C_6}(1,0,u)+\phase_{C_6}(1,0,v)=\pi$, 
therefore $p_7+p_8+p_9=1$.
Also under $C_6 \mirror C_6$ it becomes its inverse, then
$\phase_{C_6}(1,-1,v)+\phase_{C_6}(0,0,u)
+\phase_{\mirror}(0,1,u)+\phase_{\mirror}(0,0,v)
+\phase_{C_6}(0,0,v)+\phase_{C_6}(0,0,u)=\pi$,
therefore
$p_1+p_5+p_7=1$.
These constaints require $p_5=p_1$, $p_7=1$, $p_8=p_9\mod 2$. 

All nearest-neighbor ansatz on the lattice are
\begin{subequations}
\bea
A_{(x,y,u)\to(x,y,v)} & = & +A_1, \\ 
A_{(x,y,u)\to(x+1,y-1,v)} & = & +(-1)^{p_1 y}(-1)^{p_1} A_1, \\
A_{(x,y,u)\to(x,y-1,v)} & = & +A_1.
\eea
\end{subequations}

Next-nearest-neighbor ansatz $A_{<<ij>>}$: 
Assume 2nd neighbor $A_{(0,0,u)-(0,1,u)}$ is nonzero $A_2$. 
This bond under $\mirror C_6$ becomes its inverse, 
then
$\phase_{\mirror}(0,0,u)+\phase_{\mirror}(0,1,u)
+\phase_{C_6}(1,-1,v)+\phase_{C_6}(0,0,v)=\pi$,
therefore
$p_1+p_5+p_8=1$. 
Combined with constraints from nonzero nearest-neighbor ansatz,
this gives
$p_5=p_1$, $p_7=p_8=p_9=1$. 
So there is only one free Z$_2$ integer $p_1$. 

All next-nearest-neighbor ansatz on the lattice are
\begin{subequations}
\bea
A_{(x,y,u)\to(x,y+1,u)} & = & +A_2, \\
A_{(x,y,v)\to(x+1,y,v)} & = & -(-1)^{p_1 y}A_2, \\
A_{(x,y+1,u)\to(x+1,y,u)} & = & +(-1)^{p_1 y}A_2, \\
A_{(x+1,y,v)\to(x+1,y-1,v)} & = & -A_2, \\
A_{(x+1,y,u)\to(x,y,u)} & = & +(-1)^{p_1 y}(-1)^{p_1}A_2, \\
A_{(x+1,y-1,v)\to(x,y,v)} & = & -(-1)^{p_1 y}(-1)^{p_1}A_2.
\eea
\end{subequations}

With both nearest- and next-nearest-neighbor ansatz nonzero, 
there are only one free Z$_2$ integer $p_1=0,1$ in the PSG solution, 
so there are only two different Schwinger mean field theories.
The ansatz are pictorially shown in \Fig{fig:zeroflux} and \Fig{fig:piflux}.  
They are named as the zero-flux($p_1=0$) and $\pi$-flux($p_1=1$) states for 
their different gauge invariant flux in a hexagon.

Third neighbor ansatz: 
Assume third neighbor $A_{(1,-1,v)-(0,1,u)}$ is nonzero $A_3$. 
This bond under $\mirror$ becomes its inverse, 
then 
$\phase_{\mirror}(1,-1,v)+\phase_{\mirror}(0,1,u)=\pi$, 
therefore $p_7=1$. 
Also under $C_6^3$ it becomes its inverse, 
then 
$\phase_{C_6}(1,0,u)+\phase_{C_6}(0,0,v)+\phase_{C_6}(1,0,v)
+\phase_{C_6}(0,0,u)+\phase_{C_6}(1,-1,v)+\phase_{C_6}(0,1,u)
=\pi$, 
therefore $p_1+p_7+p_8+p_9=1$.
Then $A_3$ can be nonzero only in the zero-flux state ($p_1=0$). 

In the zero-flux state, all third neighbor ansatz on the lattice are
\begin{subequations}
\bea
A_{(x+1,y-1,v)\to(x,y+1,u)} & = & +A_3,\\
A_{(x+1,y,v)\to(x,y,u)} & = & +A_3,\\
A_{(x,y,v)\to(x+1,y,u)} & = & +A_3.
\eea
\end{subequations}

Fourth neighbor ansatz:
Assume 4th neighbor $A_{(0,0,v)\to(1,1,u)}$ is nonzero $A_4$.
This bond under $T_2 C_6^3$ becomes its inverse, 
then
$\phase_{C_6}(0,0,u)+\phase_{C_6}(0,1,v)
+\phase_{C_6}(1,-1,v)+\phase_{C_6}(-1,1,u)
+\phase_{C_6}(1,0,u)+\phase_{C_6}(0,-1,v)
+\phase_{T_2}(1,1,u)+\phase_{T_2}(0,0,v)=\pi$, 
therefore
$p_7+p_8+p_9=1$.
This constraint is already required by nonzero nearest-neighbor ansatz.

All fourth neighbor ansatz on the lattice are
\begin{subequations}
\bea
A_{(x,y,v)\to(x+1,y+1,u)} & = & +(-1)^{p_1 y}A_4, \\
A_{(x,y,v)\to(x,y-1,u)} & = & +A_4, \\
A_{(x,y,v)\to(x-2,y+2,u)} & = & +(-1)^{p_1}A_4, \\
A_{(x,y,v)\to(x-2,y+1,u)} & = & +(-1)^{p_1}A_4, \\
A_{(x,y,v)\to(x,y+2,u)} & = & + A_4, \\
A_{(x,y,v)\to(x+1,y-1,u)} & = & +(-1)^{p_1 y} A_4.
\eea
\end{subequations}

The PSG will also impose constraints on the $B_{ij}$ terms in \Equ{equ:SBMF}. 
For an example we consider nearest-neighbor $B_{<ij>}$. 
Assume $B_{(0,0,u)\to(0,0,v)}$ is nonzero $B_1$. 
This bond under $\mirror$ becomes its inverse $(0,0,v)\to(0,0,u)$, 
then
 $\exp\{\im[\phase_{\mirror}(0,0,v)-\phase_{\mirror}(0,0,u)]\}B_1=
(-1)^{p_9}B_1=-B_1=B_1^*$,
therefore the argument $\mathrm{Arg}(B_1^*/B_1)=\pi\mod 2\pi$. 
This bond under $T_1^{-1} C_6^3$ becomes its inverse as well, 
then 
$\phase_{C_6}(1,-1,v)-\phase_{C_6}(0,0,u)
+\phase_{C_6}(1,0,u)-\phase_{C_6}(1,-1,v)
+\phase_{C_6}(1,0,v)-\phase_{C_6}(1,0,u)
=0=\mathrm{Arg}(B_1^*/B_1)$. 
Also under $C_6 \mirror C_6$ it becomes its inverse, then
$\phase_{C_6}(1,-1,v)-\phase_{C_6}(0,0,u)
+\phase_{\mirror}(0,1,u)-\phase_{\mirror}(0,0,v)
+\phase_{C_6}(0,0,v)-\phase_{C_6}(0,0,u)=0
=\mathrm{Arg}(B_1^*/B_1)$. 
These conditions imply that $B_1$ must be zero. 

Also consider next-nearest-neighbor $B_{<<ij>>}$. 
Assume next-nearest-neighbor $B_{(0,0,u)-(0,1,u)}$ is nonzero $B_2$. 
This bond under $\mirror C_6$ becomes its inverse, 
then
$
\phase_{C_6}(1,-1,v)-\phase_{C_6}(0,0,v)
+\phase_{\mirror}(0,1,u)-\phase_{\mirror}(0,0,u)
=0=\mathrm{Arg}(B_2^*/B_2)$,
therefore 
$B_2$ must be real.

All next-nearest-neighbor $B_{<<ij>>}$ are
\begin{subequations}
\bea
B_{(x,y,u)\to(x,y+1,u)} & = & +B_2, \label{equ:B2-1}\\
B_{(x,y,v)\to(x+1,y,v)} & = & +(-1)^{p_1 y}B_2, \label{equ:B2-2}\\
B_{(x,y+1,u)\to(x+1,y,u)} & = & +(-1)^{p_1 y}B_2, \label{equ:B2-3}\\
B_{(x+1,y,v)\to(x+1,y-1,v)} & = & + B_2, \label{equ:B2-4}\\
B_{(x+1,y,u)\to(x,y,u)} & = & +(-1)^{p_1 y}(-1)^{p_1}B_2, \label{equ:B2-5}\\
B_{(x+1,y-1,v)\to(x,y,v)} & = & +(-1)^{p_1 y}(-1)^{p_1}B_2.\label{equ:B2-6}
\eea
\end{subequations}

\section{Derivation of the Continuum Field Theory
for the Transition from Zero-flux Z$_2$ Spin Liquid to
N\'eel Order}
\label{app:fieldtheory}
In this Appendix we follow the prescription of 
Sachdev\cite{SachdevEffectiveTheory} to derive the continuum field theory from 
the zero-flux Schwinger boson mean field Hamiltonian \Equ{equ:zeroflux-MF} 
close to the transition to N\'eel order. 
The notations are slightly different from Ref.~\cite{SachdevEffectiveTheory}. 
And for simplicity we omit the compact U(1) gauge field in the derivation, 
which can be added in the final result by promoting the spatial-temporal 
derivatives to covariant derivatives. 

Rewrite the bosons in terms of the would-be condensate modes 
$\psi$ at the condensation memontum $\vk=0$,  
\be
\begin{split}
&
b_{(x,y,u)\alpha}=\psi_{u\alpha}(x\veca_1+y\veca_2),
\\ &
b_{(x,y,v)\alpha}=\im\sum_{\beta}\sigma^y_{\alpha\beta}
 \psi_{v\beta}^*(x\veca_1+y\veca_2+\veca)
\end{split}
\ee
where $\vece_1=(2\veca_2-\veca_1)/3$ is the displacement of $v$ site relative 
to the $u$ site in the same unit cell. 

A gradient expansion is then performed on the real space terms in 
the mean field Hamiltonian \Equ{equ:SBMF}.
The bipartite mean field couplings become, 
up to cubic power of spatial derivatives 
(sum over spin indices $\alpha,\beta$ is implicitly assumed), 
\be
\begin{split}
& 
b_{(x,y,u)\us}b_{(x',y',v)\ds}-
b_{(x,y,u)\ds}b_{(x',y',v)\us}
\\
=\ &
-\psi_{u \alpha}
[1+
\Delta\vecr\cdot \partial_{\vecr}
+\frac{(\Delta\vecr\cdot \partial_{\vecr})^2}{2}
+\frac{(\Delta\vecr\cdot \partial_{\vecr})^3}{6}
]
\psi_{v\alpha}^*(\vecr),
\end{split}
\ee
where 
$\Delta\vecr=(x'\veca_1+y'\veca_2+\vece_1)-(x\veca_1+y\veca_2)$. 
The non-bipartite mean field couplings are 
\be
\begin{split}
&
b_{(x,y,u)\us}b_{(x',y',u)\ds}-
b_{(x,y,u)\ds}b_{(x',y',u)\us}
\\
=\ & 
\im\sigma^y_{\alpha\beta}\psi_{u\alpha}
[
\Delta\vecr\cdot \partial_{\vecr}
+\frac{(\Delta\vecr\cdot \partial_{\vecr})^2}{2}
+\frac{(\Delta\vecr\cdot \partial_{\vecr})^3}{6}
]
\psi_{u\beta},
\end{split}
\ee
and 
\be
\begin{split}
&
b_{(x,y,v)\us}b_{(x',y',v)\ds}-
b_{(x,y,v)\ds}b_{(x',y',v)\us}
\\
=\ &
\im\sigma^y_{\alpha\beta}\psi_{v\alpha}^*
[
\Delta\vecr\cdot \partial_{\vecr}
+\frac{(\Delta\vecr\cdot \partial_{\vecr})^2}{2}
+\frac{(\Delta\vecr\cdot \partial_{\vecr})^3}{6}
]
\psi_{v\beta}^*
\end{split}
\ee
where 
$\Delta\vecr=(x'\veca_1+y'\veca_2)-(x\veca_1+y\veca_2)$. 

Plug these relations into \Equ{equ:SBMF} and use the zero-flux ansatz 
\Fig{fig:zeroflux}
with nearest-neighbor and next-nearest-neighbor couplings $A_1>0$ and $A_2$. 
After collecting terms up to cubic power of spatial derivatives, 
the continuum limit Lagrangian $\mathcal{L}$ becomes 
\be
\begin{split}
&
\mathcal{L}=
\int \frac{\dif^2 \vecr}{\sqrt{3}a^2/2}\Big\{ 
\psi_{u\alpha}^* 
 \frac{\dif\phantom{\tau}}{\dif\tau}
\psi_{u\alpha} 
-
\psi_{v\alpha}^* 
 \frac{\dif\phantom{\tau}}{\dif\tau}
\psi_{v\alpha} 
\\ & 
+\mu (\psi_{u\alpha}^*\psi_{u\alpha}+\psi_{v\alpha}^*\psi_{v\alpha})
\\ &
+ A_1 \psi_{u\alpha}[\frac{3}{2}
 +\frac{\sum_{j=1}^{3}(\vece_j\cdot \partial_{\vecr})^2}{4}
 +\frac{\sum_{j=1}^{3}(\vece_j\cdot \partial_{\vecr})^3}{12}]
\psi_{v\alpha}^*
+{c.c.}
\\ &
+ A_2 (1/6) \im\sigma^y_{\alpha\beta} \psi_{u\alpha}[
 \sum_{j=1}^{3} (\vecd_j\cdot \partial_{\vecr})^3]\psi_{u\beta}
+{c.c.}
\\ &
+ A_2 (1/6) \im\sigma^y_{\alpha\beta} \psi_{v\alpha}^*[
 \sum_{j=1}^{3} (\vecd_j\cdot \partial_{\vecr})^3]\psi_{v\beta}^*
+{c.c.}
\Big \}
\end{split}
\label{equ:Lpsi}
\ee
where
${c.c.}$ means complex conjugate of the previous term, 
$\sqrt{3}a^2/2$ is the area of honeycomb unit cell, 
$a=|\veca_1|$ is the lattice constant;
$\vece_{1,2,3}$
are the three vectors connecting a $u$ site to its nearest-neighbor $v$ sites,
\be
\vece_1=(2\veca_2-\veca_1)/3,\ 
\vece_2=-(\veca_2+\veca_1)/3,\ 
\vece_3=(2\veca_1-\veca_2)/3,
\ee
and we also define for convenience
\be
 \vecd_1=-\veca_1,\ 
 \vecd_2=\veca_2,\ 
 \vecd_3=\veca_1-\veca_2.
\ee
Note that many terms are canceled due to the geometry, especially the 
first derivative terms from the $A_2$ term cancel because
$\sum_{j=1}^{3}(\vecd_j\cdot\partial_\vecr)=0$. 

Define two fields $z$ and $\Pi$ from linear combinations of 
 $\psi_{u}$ and $\psi_{v}$, 
\be
z_{\alpha}=(\psi_{u\alpha}+\psi_{v\alpha})/2,\quad
\Pi_{\alpha}=(\psi_{u\alpha}-\psi_{v\alpha})/2
\ee
Plug this into \Equ{equ:Lpsi},
the Lagrangian becomes 
(spin indices $\alpha,\beta$ are omitted),

\be
\begin{split}
&
\mathcal{L}=
\int \frac{\dif^2 \vecr}{\sqrt{3}a^2/2}\Big\{ 
2 z^*
 \frac{\dif\phantom{\tau}}{\dif\tau}
\Pi 
+2 \Pi^*
 \frac{\dif\phantom{\tau}}{\dif\tau}
z
\\ & \quad
+
(2\mu-3 A_1) z^* z+(2\mu+3 A_1)\Pi^* \Pi
\\ &\quad
+a^2(A_1/3) \partial_{\vecr} z^* \cdot \partial_{\vecr} z+{c.c.}
\\ &\quad
+(A_1/12)z^*[\sum_{j=1}^{3}(\vece_j\cdot \partial_{\vecr})^3]z+{c.c.}
\\ &\quad
+(A_2/3)
z^T(\im\sigma^y) [
 \sum_{j=1}^{3} (\vecd_j\cdot \partial_{\vecr})^3]z +{c.c.}
\end{split}
\ee
Note that terms involving both field $\Pi$ and spatial derivatives
are omitted, 
as they will generate terms in the effective Lagrangian of $z$ 
with fourth or higher power of spatial derivatives, and 
the following identity has been used,
\be
\sum_{i=1}^{3}(\vece_i\cdot \partial_{\vecr})^2
=(2/3)a^2\partial_{\vecr}^2
\ee

Integrate out the field $\Pi$ with large gap $2\mu+3 A_1$,  
we get the effective Lagrangian for $z$
\be
\begin{split}
& 
\mathcal{L}_{z}=
\int \dif^2 \vecr \Big [
\frac{8}{(2\mu+3 A_1)\sqrt{3}a^2}\partial_{\tau}z^*\cdot \partial_{\tau}z
\\ &\quad
+\frac{2 A_1}{3\sqrt{3}} \partial_{\vecr} z^* \cdot \partial_{\vecr} z
+\frac{2(2\mu-3 A_1)}{\sqrt{3}a^2} z^* z
\\ &\quad
+\frac{A_1}{6\sqrt{3}a^2}z^*[\sum_{j=1}^{3}(\vece_j\cdot \partial_{\vecr})^3]z
+{c.c.}
\\ &\quad
+\frac{2 A_2}{3\sqrt{3}a^2}
z^T (\im\sigma^y) [
 \sum_{j=1}^{3} (\vecd_j\cdot \partial_{\vecr})^3]z+{c.c.}
\Big ]. 
\end{split}
\ee
The critical point is $A_1/\mu=2/3$ consistent with the mean field solution. 
The critical boson velocity is proportional to $A_1$. 
After a proper rescaling of $\tau$
the Lagrangian can be cast into the simple form of \Equ{equ:action}. 
Note that $A_2$ plays the role of the Higgs field.

\end{document}